\documentclass[twocolumn,aps,pra,showpacs,bibfootnote,preprintnumbers, mathtools, superscriptaddress, amsmath,amssymb]{revtex4-2}

\usepackage{graphicx}
\usepackage{dcolumn}
\usepackage{bm}
\usepackage{bbold}
\usepackage{braket}
\usepackage{xcolor}
\usepackage{changes}
\usepackage{hyperref}

\begin{document}

\title{Optical Distinguishability of Mott Insulators in Time vs Frequency Domain}


\author{Jacob Masur \href{https://orcid.org/
0000-0002-2477-6917}{\includegraphics[scale=0.05]{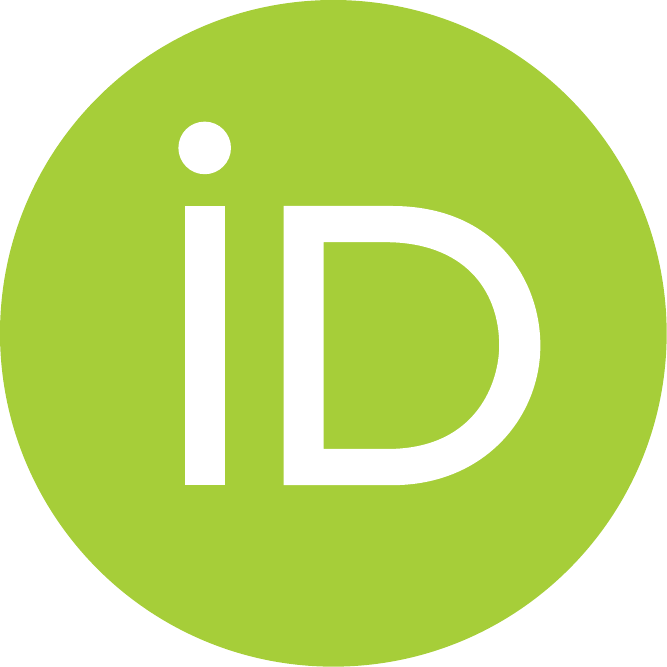}}}
\email{jmasur@tulane.edu}
\affiliation{Tulane University, New Orleans, LA 70118, USA}

\author{Denys I. Bondar \href{https://orcid.org/0000-0002-3626-4804}{\includegraphics[scale=0.05]{orcidid.pdf}}}
\affiliation{Tulane University, New Orleans, LA 70118, USA}

\author{Gerard McCaul
\href{
https://orcid.org/
0000-0001-7972-456X}{\includegraphics[scale=0.05]{orcidid.pdf}}}
\email{gmccaul@tulane.edu}
\affiliation{Tulane University, New Orleans, LA 70118, USA}

\date{\today}

\begin{abstract}
    High Harmonic Generation (HHG) promises to provide insight into ultrafast dynamics and has been at the forefront of attosecond physics since its discovery. One class of materials that demonstrate HHG are Mott insulators whose electronic properties are of great interest given their strongly-correlated nature. Here, we use the paradigmatic representation of Mott insulators, the half-filled Fermi-Hubbard model, to investigate the potential of using HHG response to distinguish these materials. We develop an analytical argument based on the Magnus expansion approximation to evolution by the Schrodinger equation that indicates decreased distinguishability of Mott insulators as lattice spacing, $a$, and the strength of the driving field, $F_0$, increase relative to the frequency, $\omega_0$. This argument is then bolstered through numerical simulations of different systems and subsequent comparison of their responses in both the time and frequency domain.  Ultimately, we demonstrate reduced resolution of Mott insulators in both domains when the dimensionless parameter $g \equiv aF_0 / \omega_0$ is large, though the time domain provides higher distinguishability. Conductors are exempted from these trends, becoming much more distinguishable in the frequency domain at high $g$.
\end{abstract}

\maketitle

\section{Introduction}

One of the principal breakthroughs of the last half-century has been the discovery of non-linear optics, enabled by the invention of the laser \cite{PhysRevLett.7.118, PhysRevLett.7.229}. Non-linear effects are expected to play a crucial role in what has been termed the `second quantum revolution' \cite{Dowling2003}, where quantum effects are exploited to develop new technology. It has already been demonstrated that systems exhibiting a non-linear optical response possess a number of both useful and surprising properties, such as controllability \citep{PhysRevA.72.023416,PhysRevA.98.043429,Campos2017,Gerard1,Gerard2}, optical indistinguishability \cite{Busson2009, Gerard3}, non-uniqueness \cite{Ong1984,Clark1985,jha_multiple_2009,Gerard4}, self-focusing \cite{PhysRevLett.13.479}, the nonlinear Fano effect \cite{Kroner2008}, and many more \cite{shen1984principles, de_Ara_jo_2016}.

Within the plethora of non-linear effects already discovered \cite{boyd_nonlinear_2008}, one of the most promising is High Harmonic Generation (HHG). First observed in gases, this phenomenon has also been induced by strong fields in atoms and molecules \cite{Itatani2004, PhysRevA.65.053805, PhysRevA.65.053805, PhysRevA.66.023805, PhysRevA.67.023819, PhysRevA.70.011404, PhysRevA.71.053407,electronHHG}, the surfaces of both metal and dielectric solids \cite{PhysRevA.52.R25, Dromey2006, PhysRevLett.113.073901, RevModPhys.90.021002, ORTMANN2021103}, and in bulk crystals \cite{ObsOfHHG, Vampa2015, Ndabashimiye2016, Luu2015, Ndabashimiye2016, Hohenleutner2015, Langer2016, Yoshikawa736, Liu2017, anisoHHGinCrystals, HHGinSemiconductors}. The ability to generate light at frequencies many multiples greater than the initial excitation provides a tool for observation and manipulation at the attosecond timescale \cite{Ciappina_2017,Drescher1923, Paul1689, Hentschel2001}. Given inter- and intra-atomic electron motion occurs on precisely this time scale, applications of HHG can provide a route to the study of the electronic structure of materials \cite{Paul1689, anisoHHGinCrystals, attosecond}. Not only does this offer insight into fundamental phenomena such as  tunneling \cite{LANDSMAN20151}, proper understanding of atomic scale properties has many applications, from mixture characterisation \cite{mixing_mccaul}, to faster task-specific electronics \cite{Krausz_2001}, and measurements of chirality \cite{Cireasa2015, PhysRevX.9.031002} to novel ``valleytronics" \cite{jimenez-galan_sub-cycle_2021}.

Given that HHG offers a higher resolution window on electron dynamics, it is natural to ask how it might be used to improve material characterisation. In the modern era, this process has moved beyond the determination  of simple properties such as conductivity, and focuses instead on microscopic or even atomic properties. To do so, tools such as scanning tunneling microscopes \cite{mottMeter} and electron diffraction \cite{Chergui1002} have been employed. These have been had great success in certain applications, but are far from ubiquitous \cite{anisoHHGinCrystals}. Meanwhile, high harmonic responses to optical driving has become a useful tool in electronic analysis \cite{attosecond}. Researchers have successfully applied HHG spectroscopy to probe the electronic structure of specific materials \cite{Itatani2004, PhysRevA.65.053805, PhysRevA.66.023805, PhysRevA.67.023819, PhysRevA.70.011404, PhysRevA.71.053407, PhysRevA.52.R25, ObsOfHHG, Vampa2015, Dromey2006, Ndabashimiye2016, Luu2015, Hohenleutner2015, Langer2016, Yoshikawa736, Liu2017, anisoHHGinCrystals, PhysRevLett.88.183903, Hentschel2001, PhysRevLett.113.073901, HHGinSemiconductors, RevModPhys.90.021002, electronHHG, ORTMANN2021103}, and it has even been shown recently that HHG spectra can indicate a critical point of a quantum phase transition from a spin density wave to a charge density wave \cite{shao2021detecting}. Taken together, these results indicate that the high harmonic response of materials may be a powerful future tool in the characterisation and analysis of of materials in the solid state.

In such a case, one would ideally employ all information that can be gleaned from a material's optical emission - i.e. the full time-domain response. While there is some prospect of obtaining this in the future \cite{attosecond}, at present time-resolved measurement of subfemtosecond processes is difficult in the solid-state \cite{Krausz_2001, Garg2021}. Nevertheless, frequency-domain spectroscopy of high-order harmonics and their relative intensities can be acquired experimentally, at the cost of losing the phase information of the response. This begs the question as to the importance of this lost phase information.

Here, we consider the potential of high harmonic optical responses as a tool for material identification while placing particular emphasis on the role of the optical response's phase information in distinguishing the driven dynamics of different materials. We investigate this question in a class of materials described by the Hubbard Model, namely Mott insulators \cite{simon2013oxford, Hubbard}. Previous studies of HHG in the Hubbard Model have revealed that the mechanism for generation of high harmonics is intrinsically distinct from that of crystals and other solids in which HHG has been observed \cite{HHGinMottInsulators, PhysRevLett.124.157404}. Specifically, the plateaus in the optical spectra depend heavily on the dynamics of the charge carriers, doublons and holes \cite{HHGinMottInsulators}, suggesting that the HHG spectra of Mott insulators contain valuable electronic information \cite{Silva2018}. This naturally begs the question as to whether a specific Mott insulator can be characterised and distinguished from others purely by its high harmonic response. This would in turn facilitate the parametrisation of new materials purely by a spectral measurement.

The rest of this paper will be structured as follows. In Sec.\ref{section:Methods} we set out the the driven Hubbard model, and define figures of merit which quantify the relative distinguishability of two systems based on their time or frequency domain response. We also provide a heuristic analysis predicting how the degree of distinguishability of two materials depends on both intrinsic system and laser pulse parameters. Sec.\ref{section:Results} presents the results of numerical analysis, identifying the regime in which phase information plays an important role in material characterisation. Finally, we close with a discussion of the results and their potential applications in Sec.\ref{section:conclusion}.

\section{Model \label{section:Methods}}

We would like to analyze the responses of different materials to an incident laser pulse which couples electrons to an electric field. Using atomic units ($\hbar = e = 1$) henceforth, unless explicitly stated otherwise, we take the paradigmatic Hubbard model \cite{Hubbard, Gerard2, HHGinMottInsulators, PhysRevLett.124.157404, Silva2018}:
\begin{align}
    \hat{H}(t) = -t_0 \sum_{j,\sigma}( e^{-i\Phi(t)} \hat{c}^{\dag}_{j,\sigma} \hat{c}_{j+1,\sigma} + \mathrm{h.c.}) + U\sum_j \hat{n}_{j, \uparrow} \hat{n}_{j, \downarrow} \label{Hamiltonian}
\end{align}
where $\hat{c}^{\dag}_{j,\sigma}$ and $\hat{c}_{j,\sigma}$ are, respectively, the canonical fermionic creation and annihilation operators at site $j$ with spin $\sigma$, $\hat{n}_{j, \sigma} = \hat{c}^{\dag}_{j, \sigma} \hat{c}_{j, \sigma}$ is the particle number operator, $t_0$ is the hopping parameter, $U$ is the onsite interaction parameter,  and $\Phi(t) = aA(t)$ is the Peierls phase. This in turn is composed of both lattice spacing $a$ and the vector potential $A(t)$, which is related to the electric field by $E(t) = -dA(t)/dt$.

Though the Hubbard Model is a relatively simple description of Mott insulators, it provides rich insight into the strong electron-electron interactions in these materials \cite{Hubbard, Silva2018, simon2013oxford, AshcroftMermin}, distinguishing itself from models relying on mean field approximations. These interactions play a crucial role in the band-structure of Mott insulators \cite{HHGinMottInsulators}, preventing current flow in these materials with half-filled valence bands that are predicted to be conductors under band theory.
\begin{figure}[ht]
\begin{center}
\includegraphics[width=1\columnwidth]{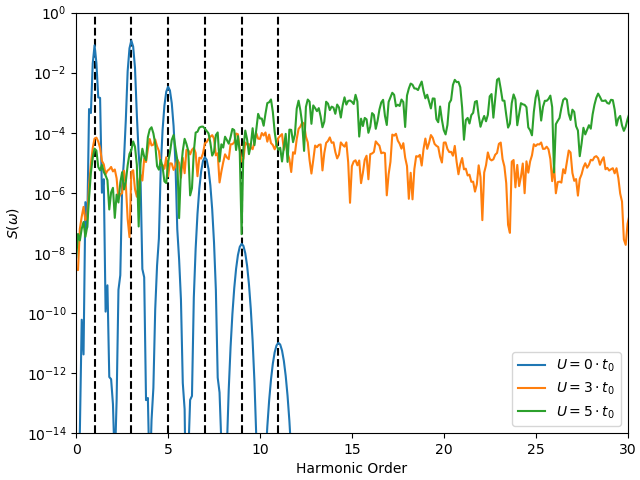}\end{center}
\caption{Power spectra of a conductor ($U/t_0 = 0$) and two Mott insultors ($U/t_0 = 3$ and $U/t_0 = 5$) at $a$ = 4\r{A} and $F_0 = 10 \frac{\textrm{MV}}{\textrm{cm}}$. The vertical lines indicate odd integer harmonics.}
\label{fig:spectra}
\end{figure}

The optical response of the material to  driving is quantified by the current operator:
\begin{align}
    \hat{J}(t) = -iat_0 \sum_{j,\sigma} e^{-i\Phi(t)} \hat{c}^{\dag}_{j,\sigma} \hat{c}_{j+1,\sigma} - {\rm h.c.} \label{CurrentOp}
\end{align}
which can be derived from a continuity equation for electron density \cite{Gerard2}. The expectation of the current density operator over the course of the evolution gives access to all spatial and temporal information regarding a given material's optical response. Ordinarily, it is easier to measure the spectra rather than the time resolved HHG output. The spectrum is given by
\begin{equation} \label{spectrum}
    S(\omega) = |\mathcal{F}_{t\rightarrow\omega} J(t) |^2, 
\end{equation}
where $\mathcal{F}$ denotes the Fourier transform and $J(t)=\langle \hat{J}(t) \rangle$ is the current expectation. 

Clearly, the spectrum does not contain any of the phase information present in the corresponding current expectation.  We seek to determine not only the degree to which  one is able to distinguish between systems via their optical response, but whether phase information materially affects this distinguishability. In order to assess the relative distinguishability of two systems (indexed by $i$ and $j$), we define two relative distance functions for a set of current expectations and their spectra:
\begin{align}
&\mathcal{D}_t (i,j) =  \sqrt{\int_0^T\left( \frac{J_i(t)}{\bar{J}_i} - \frac{J_j(t)}{\bar{J}_j} \right)^2 {\rm d}t}, \label{timecost} \\
&\mathcal{D}_p (i,j) = \sqrt{\int_{0}^{\omega_c}\left( \log\left[\frac{S_i(\omega)}{\bar{S}_i}\right]- \log\left[\frac{S_j(\omega)}{\bar{S}_j}\right] \right)^2 {\rm d}\omega} \label{spectrumcost}, \\
&\bar{J}_i=\max_{t\in[0,T]} J_i(t), \ \ \ \ \ \ \ \ \ \ \ \ \  \bar{S}_i=\max_{\omega\in[0,\omega_c]} S_i(\omega).
\end{align}

Here $\mathcal{D}_t$ denotes the distance in the time domain, $\mathcal{D}_p$ denotes the distance in the frequency domain. $T$ is the duration of the pulse, and $\omega_c$ is the cutoff frequency defined as the minimum frequency for which $S_i(\omega_c) = 10^{-20}$ or $S_j(\omega_c) = 10^{-20}$, choosing whichever $\omega_c$ is smaller. Note that these expectations are scaled by their maximum value over the pulse, reflecting the fact that Hubbard dynamics are controlled by the ratio $\frac{U}{t_0}$ \cite{Hubbard}, rather than any absolute scaling of the Hamiltonian. This ensures that the distinguishability measures capture genuine differences in the optical dynamics, rather than simple scale variations. Naturally, a distance of zero between two systems indicates identical responses and perfect indistinguishability, while a larger distance indicates a higher degree of relative distinguishability. 

In equilibrium ($\Phi(t)=0$) only the ratio of two parameters, the interaction energy ($U$) and the hopping energy ($t_0$), determine the electronic properties of a material described by Eq. \eqref{Hamiltonian}. Consequently, one may scale all parameters to units of $t_0$,  and remove it from consideration. Under driving however, the dynamics of the optical response depends not only on $\frac{U}{t_0}$, but the lattice constant $a$ and the driving pulse. For concreteness, we consider driving each system from its ground state with the phase resulting from a transform-limited laser pulse:
\begin{equation} \label{inputfield}
    \Phi(t) = g \sin^2(\frac{\omega_0 t}{2M})\sin(\omega_0 t)
\end{equation}
where $M = 10$ is the number of cycles, and the dimensionless ratio $g = aF_0/\omega_0$ relates the lattice constant, driving frequency $\omega_0$, and field strength $F_0$. It is interesting to note the similarity between $g$ and the Hubbard model Keldysh crossover parameter $\gamma = \omega_0/(\xi F_0)$ \cite{Oka, keldysh1965ionization}, where $\xi$ is the correlation length \cite{millis0}. This parameter indicates the mechanism for pair production: $\gamma \ll 1$ defines the multiphoton absorption regime whereas $\gamma \gg 1$ defines the quantum tunneling regime \cite{Oka}.

The response of a given material to optical stimulation depends heavily on the values of $U$ and $g$ that define the material's electronic properties. For Mott insulators, charge is carried by doublon-hole excitations \cite{HHGinMottInsulators}, the pair production of which decreases exponentially with $U$ \cite{Oka}. Fig. \ref{fig:spectra} show some of the essential optical characteristics resulting from this, such as a white light response to optical driving in sufficiently strong Mott insulators. In the $U=0$ conducting limit, the spectral behaviour becomes identical to atomic systems \cite{PhysRevLett.68.3535}, with well defined peaks present at odd harmonics. 

There is no simple method for determining the effect of $g$ on material response, but we can obtain a heuristic understanding of how this factor affects optical spectra in the limit of large $g$. Taking $|\psi(t)\rangle = \hat{U}(t)|\psi(0)\rangle$, the propagator will be of the form:
\begin{equation} \label{propagatorsolution}
    \hat{U}(t) = \exp[\Omega(t)]
\end{equation}

We would like to approximate an effective time-independent Hamiltonian, such that:
\begin{equation} \label{effham}
\Omega(t) \approx -i\hat{H}_{\rm eff} t
\end{equation}
$\Omega(t)$ can be expressed in terms of the Magnus expansion \cite{magnus}:
\begin{equation} \label{expansion}
    \Omega(t) = \sum_{k=1}^{\infty} \Omega_k (t).
\end{equation}
The first term in this expansion is given by
\begin{align}
    \Omega_1(t) ={}& -i \int_0^t \hat{H}(t_1) {\rm d}t_1 \label{term1}.
\end{align}
The integral over the two body term of Eq.\eqref{Hamiltonian} is trivial given that it is time independent. Thus, we focus our attention on the integral over the hopping term:
\begin{align}
   &\int_0^T {\rm d}t\sum_{j,\sigma} e^{-i\Phi(t)} \hat{c}^\dag_{j, \sigma}\hat{c}_{j+1, \sigma} + {\rm h.c}. \\
    = &\sum_{j,\sigma}\hat{c}^\dag_{j, \sigma}\hat{c}_{j+1, \sigma} \int_0^T {\rm d}t e^{-i\Phi(t)}  + {\rm h.c.}
\end{align}
where integrating over the duration of the single pulse, $T = 2\pi M/ \omega_0$. The relevant integral is therefore
\begin{equation} \label{integral}
    I = \int_0^T {\rm d} t\ \exp\left[-i g f(t)\right] 
\end{equation}
where $f(t) = \sin^2(\omega_0 t/ 2M)\sin(\omega_0 t)$.

In order to restrict the analysis to this first order term, we treat the scaling factor $g$ as a large parameter, meaning higher order terms in the Magnus expansion will oscillate rapidly and can be neglected.  Given the envelope is slowly varying relative to $\omega_0$, for each half period $\frac{\pi}{\omega_0}$, we approximate it by its averaged value over that period:
\begin{equation}
\Delta_j= \frac{1}{2}\left[\sin^2\left(\frac{j\pi}{2M}\right)+\sin^2\left(\frac{(j+1)\pi}{2M}\right)\right].    
\end{equation}
\begin{figure}[ht]
\begin{center}
\includegraphics[width=1\columnwidth]{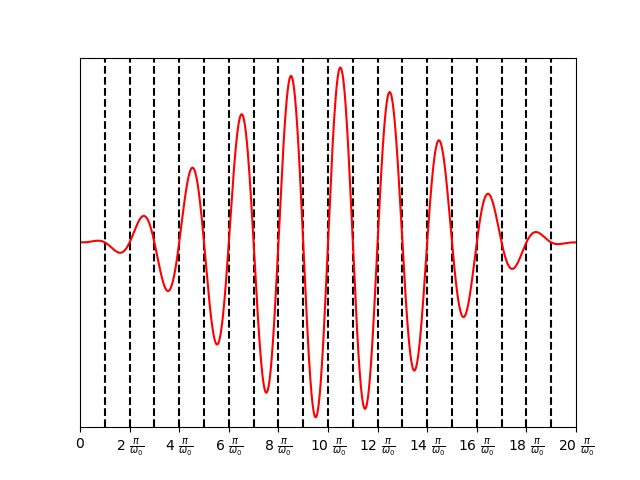}\end{center}
\caption{A plot of $f(t)$ where each dashed line is the boundary of a $\Delta_j$.}
\label{fig:pulse}
\end{figure}
Consequently, the integral can be approximated as
\begin{align}
    I \approx& \frac{2}{\omega_0}\sum^{M-1}_{j=0} \int_{0}^{\pi} {\rm d}t\ \cos\left[g\Delta_j\sin(t)\right] \notag \\
    & =\frac{2}{\omega_0}\sum^{M-1}_{j=0}  J_0\left(g\Delta_j\right).
\end{align}
where  $J_0 \left(x\right)$ is the zero order Bessel function \cite[Eq.~(10.9.1)]{NIST:DLMF}. In the limit of large $g$, this function may be replaced by its asymptote \cite[Eq.~(10.17.3)]{NIST:DLMF} to obtain
\begin{equation}
    I \approx T \sqrt{\frac{1}{g}} \zeta,
\end{equation}
where 
\begin{equation}
    \zeta = \frac{2}{T}\sqrt{\frac{2}{\pi}}\sum^{j=M-1}_{j=0} \frac{1}{\sqrt{\Delta_j}}\cos\left(g\Delta_j-\frac{\pi}{4}\right).
\end{equation}

Having approximated the integral, it is now possible to state  the form of the effective Hamiltonian in Eq. \eqref{effham}:
\begin{equation}
    \hat{H}_{\rm eff} \approx -\frac{\zeta t_0}{\sqrt{g}} \sum_{j,\sigma} \left(\hat{c}^\dag_{j, \sigma}\hat{c}_{j+1, \sigma} + {\rm h.c.}\right) + U\sum_j \hat{n}_{j,\uparrow}\hat{n}_{j,\downarrow}.
\end{equation}
This effective Hamiltonian is in the form of the Hubbard Hamiltonian in equilibrium with one important difference, the scaling of the hopping energy by $\frac{\zeta}{g}$. 

\begin{figure}[ht]
\begin{center}
\includegraphics[width=\columnwidth]{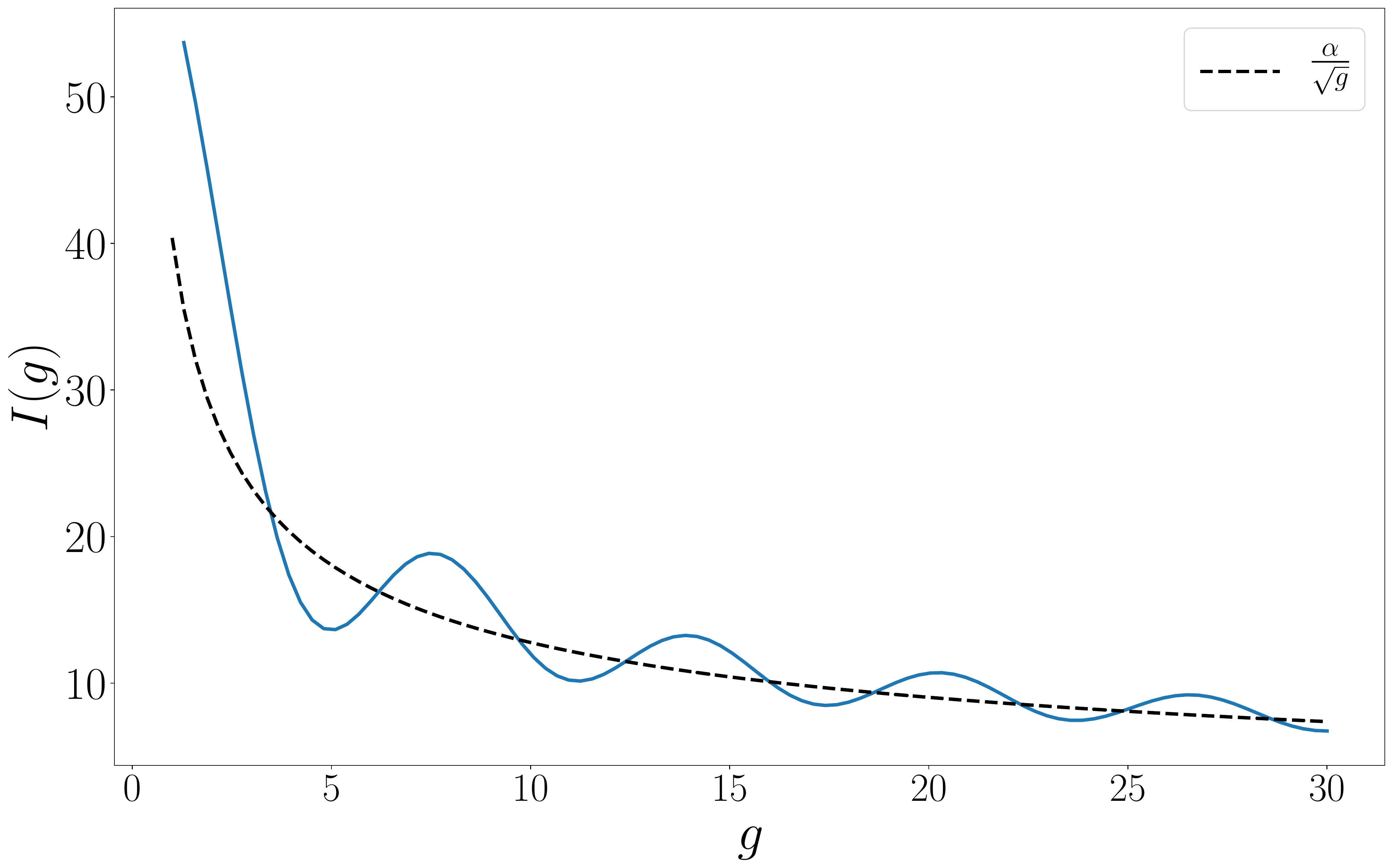}\end{center}
\caption{Numerical calculation of Eq.\eqref{integral}. As expected, the behaviour at large $g$ is well described by $\frac{\alpha}{\sqrt{g}}$, where $\alpha$ is some constant of proportionality. The trigonometric dependence of $\zeta$ on $g$ introduces small oscillations around the central asymptote.}
\label{fig:integral_approx}    
\end{figure}

Note that while $\zeta$ has some $g$ dependence, this occurs only in the trigonometric terms in the sum, meaning that this parameter will be bounded with respect to $g$, and have little impact once the response is normalised. This is borne out by numerical calculation of Eq.\eqref{integral}, as shown in Fig. \ref{fig:integral_approx}. Here it is apparent that the contribution of $\zeta$ at large $g$ is the introduction of small oscillations around $\frac{1}{\sqrt{g}}$. It therefore follows that the principal effect of increasing $g$ will be the scaling of the system response in the following manner:
\begin{equation} \label{scaling}
    t_0 \rightarrow \frac{t_0}{\sqrt{g}} \Longrightarrow \frac{U}{t_0} \rightarrow \frac{U\sqrt{g}}{t_0}.
\end{equation}

\begin{figure}[ht]
\begin{center}
\includegraphics[width=\columnwidth]{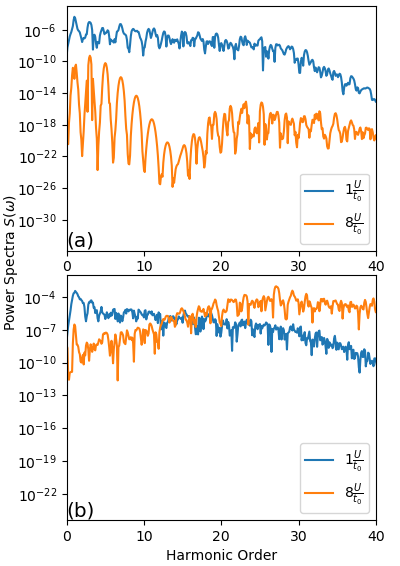}\end{center}
\caption{Spectra of two systems defined by their intrinsic electronic properties ($\frac{U}{t_0} = 1$ and $\frac{U}{t_0} = 8$) at two levels of scaling: the spectra in plot are scaled by $g = 4.62$ and the spectra in plot (b) are scaled by $g = 37.0$.}
\label{fig:scaled_spectra}    
\end{figure}

Clearly, for any insulator, the effective ratio of the interaction parameter to the hopping parameter is proportional to the scaling parameter, so an increase in $g$ leads to a higher effective ratio, one where $U$ dominates the dynamics of the system. Consequently, increasing $g$ means all systems are shifted towards the high $U$, strongly insulating regime. The one exception to this is the $U=0$ conducting system, which will only experience a scaling in the magnitude of its optical response, rather than its dynamical character. 

Thus, when $g$ is increased, we expect that the distance between the responses of two Mott insulators given by Eqs. \eqref{timecost} and \eqref{spectrumcost} to reduce, regardless of the specific ratio corresponding to each system. This is exactly the phenomena illustrated in Fig. \ref{fig:scaled_spectra}: at low scaling the two systems have highly distinguishable responses, but at high scaling their power spectra are more similar. The only type of material that will be distinguishable from Mott insulators in the high $g$ regime are conductors whose response is unaffected by the scaling in Eq. \eqref{scaling}.

\begin{figure*}
  \includegraphics[width=.9\textwidth]{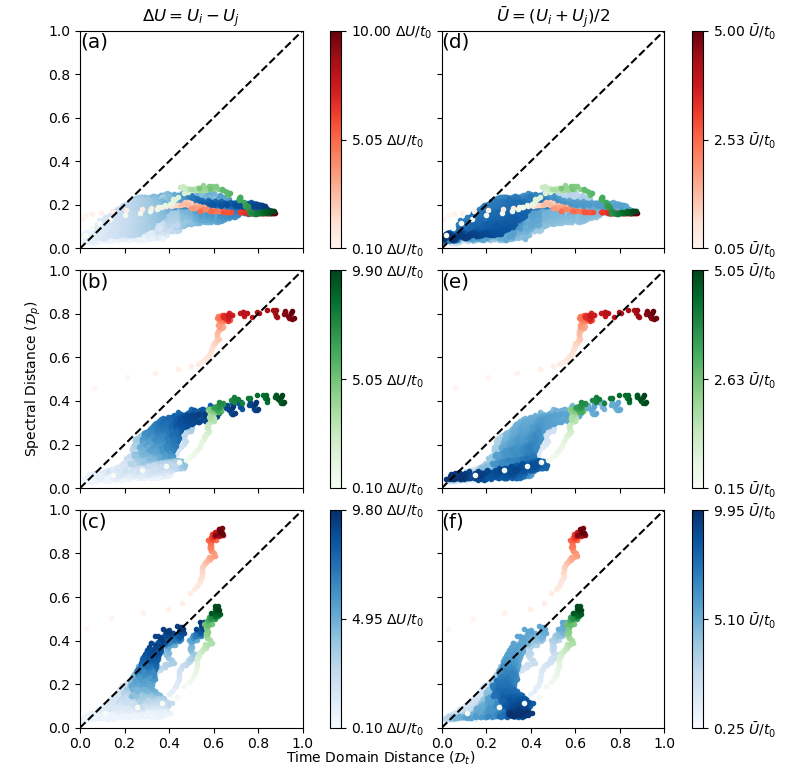}
  \caption{Each point is a comparisons of two systems with different electronic properties ($U_i \neq U_j$) at the same scaling factor ($g$). Systems represented in plots (a) and (d) are driven with $g = 4.62$, those in plots (b) and (e) are driven with $g= 14.7$, and those in plots (c) and (f) are driven with $g = 25.2$. The coloring of the points in plots (a) - (c) and (d) - (f) represents the $\Delta U = U_i - U_j$ and $\bar{U} = (U_i + U_j) / 2$ of the systems being compared, respectively. A more saturated color indicates a larger difference/average of the $U$ values, and a less saturated color indicates a smaller difference/average. The different coloring schemes highlight special comparisons. Red points in plots (a) - (c) and orange points in plots (d) - (f) are comparisons where one system is a conductor ($U_i = 0$), green points in plots (a) - (c) and grey points in plots (d) - (f) are comparisons to the smallest nonzero value of $U_i$, $U_1 = 0.1 t_0$, and the blue points in plots (a) - (c) and purple points in (d) - (f) are all remaining comparisons, that is, comparisons of only insulators.}
  \label{fig:scatter_diff}
\end{figure*}

\section{Results \label{section:Results}}

In this section, we numerically investigate the extent to which the degree of distinguishability between systems depends on both system $\frac{U}{t_0}$, $a$ and pulse parameters $F_0$ and $\omega_0$. In particular, we demonstrate that there is a comparatively high experimental optical discriminability between many Mott insulators at a low lattice spacing and low field strength relative to the driving frequency. Conversely, as predicted in Sec.\ref{section:Methods}, as one increases the factor $g$, all systems (except conductors) exhibit behaviour associated with the large $\frac{U}{t_0}$ regime, and hence become less distinguishable. The increased distinguishability of conductors in this parameter range is also demonstrated.

Pairs of systems distinguished by potentials $U_1$ and $U_2$ are simulated via exact diagonalization in QuSpin \cite{quspin}. From this, the distance measures of Eqs.(\ref{timecost},\ref{spectrumcost}) are calculated. These distances are then scaled to their maximum value over the total range of parameter pairs considered, to give a measure of relative distinguishability. Fig. \ref{fig:scatter_diff} summarises these results, reporting relative distinguishability in the time and spectral domains via both the difference $\Delta U$ and average $\bar{U}$ of each system's potential.

The first point to note is that in almost all cases, the relative distinguishability of two systems is greater in the time domain than the spectral domain (i.e. points lie below the diagonal), demonstrating the importance of phase information for distinguishing between system responses. As might be expected, distinguishability tends to be lowest for materials with small $\Delta U$ and high $\bar{U}$. The relative distinguishability between the two domains also changes dramatically as $g$ is increased. 

The clearest example of this is in the behaviour of system pairs where one system is a conductor ($U=0$). In this case, while the time domain distinguishability is reduced with higher field amplitudes, the spectral distinguishability increases significantly, such that at high driving fields spectral characteristics provide a greater degree of distinguishability compared to the time domain.  

In fact, when examining both difference and average, a clear trend that emerges where systems close to the conducting limit increase in spectral distinguishability as $g$ increases. This is to be expected, given the scaling argument in Eq. \eqref{scaling}. Since the effective ratio of conductors cannot be scaled by the field strength or lattice constant, conductor response to optical driving remains is only scaled, rather than dynamically changed by varying $g$. A system close to this $U=0$ limit will still retain its conductor-like properties, whereas a system that is already deeply in the insulating regime is scaled into an even more strongly insulating system. This behaviour is most strikingly observed in Fig. \ref{fig:scatter_diff} c), where at high $g$ a banding effect is observed separating distinguishabilities into those systems where one of the material pair is either a conductor or small $U$ material. 

Finally, we find in all cases that while increasing $g$ may increase spectral distinguishability, for almost all pairs in the time domain, a steady compression of distinguishability along this axis can be observed, as one might expect from Eq.\eqref{scaling}.  This is itself strongly dependent on $\bar{U}$, with a relatively small subset of pairs featuring both high $\Delta U$ and small $\bar{U}$ becoming relatively more distinguishable in the spectral domain than the time domain at high $aF_0$ (i.e. Fig. \ref{fig:scatter_diff} c) and f)).

\section{Discussion \label{section:conclusion}}

Here we have examined the feasibility of distinguishing driven Mott insulators via their optical response. Simulation demonstrated that the importance of phase information in this process was dependent on the applied field strength, and confirmed the heuristic argument that the distinguishability of insulators should decrease as the field strength is increased. To paraphrase Tolstoy \cite{tolstoy2013anna}, conductors - like happy families - always retain a high degree of distinguishability in one domain or another, whereas the distinguishability of insulating pairs depends strongly on the driving field amplitude. 

The high dependence of material response on the scaling factor $g$ begs the question, what scaling values are physically realizable for  high frequency pulses? Considering a simple subclass of Mott insulators, transition metal oxides, we determine that 4 \r{A} is physically realistic lattice constant based on studies performed by Heine and Mattheiss \cite{Heine_1971}. Hohenleutner et al. \cite{Hohenleutner2015} also experimentally demonstrated pulse generation with a peak field of 44 $\frac{\mathrm{MV}}{\mathrm{cm}}$. Thus, given an infrared frequency $32.9$ THz, the scaling factors studied here are certainly within the range of allowed experimental values, since these estimates place a maximum scaling at $g = 81.3$. Moreover, the results presented here demonstrate that the change in distinguishability sweeping over a range of field strengths may also serve as a useful source of information for identifying materials. 

It is rather interesting to note that the strong dependence of distinguishability on $g$ implies that the Keldysh parameter $\gamma$ may \emph{also} serve as a proxy for optical distinguishability for systems. Indeed, given this parameter will depend on each system's correlation length (and therefore $U$), it encapsulates more information about each system individually than $g$ alone. A potential future avenue of investigation would  there be to study whether a pairwise function of each system's $\gamma$ may offer some predictive heuristic for their relative distinguishability.

Of particular importance is the finding that experimental differentiation between different Mott insulators is in most cases more easily achieved in the time domain. While spectral characteristics are unquestionably easier to obtain experimentally, the results presented show that the technique of THz time domain spectroscopy \cite{TDS2,TDS1} could profitably be employed to better distinguish between materials.

Though our results only apply to strongly-correlated materials, sufficient experimental optical similarity between a known Mott insulator whose atoms are spaced sufficiently and a material with comparable interatomic spacing could provide an aid to the nontrivial problem of distinguishing between a band insulator and a Mott insulator \cite{mottMeter}. 

\begin{acknowledgements}
This work has been generously supported by Army Research Office (ARO) (grant W911NF-19-1-0377; program manager Dr.~James Joseph). J.M. was supported by the ARO Undergraduate Research Apprenticeship Program and the Tulane Honors Summer Research Program. The views and conclusions contained in this document are those of the authors and should not be interpreted as representing the official policies, either expressed or implied, of ARO or the U.S. Government. The U.S. Government is authorized to reproduce and distribute reprints for Government purposes notwithstanding any copyright notation herein.
\end{acknowledgements}

\bibliography{bibliography}

\end{document}